\documentclass[twocolumn]{article} 
\usepackage{xspace}
\usepackage{url}
\usepackage{mathtools}
\usepackage{nicefrac}
\usepackage{multirow}
\usepackage{amsmath,amsthm,amsfonts}

\usepackage{balance}
\usepackage{paralist}
\newcommand{\mysub}[1]{\smallskip\noindent\textbf{#1}:}
\newcommand{\mysubb}[1]{\smallskip\noindent\textbf{#1}}
\newcommand{\mypara}[1]{\smallskip\noindent\emph{#1}:\vspace{-0.1cm}}
\usepackage{enumitem}
\setlist[itemize]{leftmargin=*}
\setlist[enumerate]{leftmargin=*}
\title{Foundations for Grassroots Democratic Metaverse}

\author{
  Ehud Shapiro\\
  Weizmann Institute of Science\\ Columbia University\\
  \texttt{ehud.shapiro@weizmann.ac.il}
  \and
  Nimrod Talmon\\
  Ben-Gurion University\\
  \texttt{talmonn@bgu.ac.il}
}

\date{}

\begin{document}
\pagestyle{plain}
\maketitle

\begin{abstract}
While the physical lives of many of us are in democracies (one person, one vote — e.g., the EU and the US), our digital lives are mostly in autocracies (one person, all votes — e.g., Facebook). Cryptocurrencies promise liberation but stop short, at plutocracy (one coin, one vote). What would it take for us to live our digital lives in a digital democracy?  This paper offers a vision, a theoretical framework, and an architecture for a grassroots network of autonomous, people-owned,  people-operated, and people-governed digital communities, namely a \emph{grassroots democratic metaverse}.  It also charts a roadmap towards realizing it, and identifies unexplored territory for further research.
\end{abstract}


\section{Introduction}

A key obstacle to digital equality and digital democracy is fake and duplicate digital identities, aka \emph{sybils}~\cite{douceur2002sybil}. Facebook eliminates billions of sybils every quarter~\cite{facebook-sybils} and even if it would decide to go democratic~\cite{facebookdictatorship}, it is technically unable to~\cite{facebooksybils}. Cryptocurrencies employ plutocratic proof-of-work~\cite{bitcoin} or proof-of-stake~\cite{kiayias2017ouroboros}, partly for lack of a better way to defend against sybils~\cite{de2019blockchain}.

Here we describe a vision, a theoretical framework, and an architecture for the grassroots formation of a network of autonomous, people-owned,  people-operated, and people-governed digital communities, referred to as  \emph{democratic DAOs} (Decentralized Autonomous Organizations), which jointly form a \emph{grassroots democratic metaverse}.  
A central theme of our approach is \emph{sybil resilience}: means to minimize sybil penetration into the democratic digital communities that form the metaverse, and means for a digital democracy to function despite a limited sybil penetration.
More concretely, our proposed architecture for a grassroots democratic metaverse calls for the design and implementation of several inter-related components, which have equality as a fundamental tenet, and to achieve that must be sybil-resilient.

Figure \ref{figure:metaverse} presents the integrated architecture we envision, with the following components (in each corresponding section of the manuscript, we highlight specific challenges for MAS research):
\begin{figure*}[t]
\centering
\includegraphics[width=15cm]{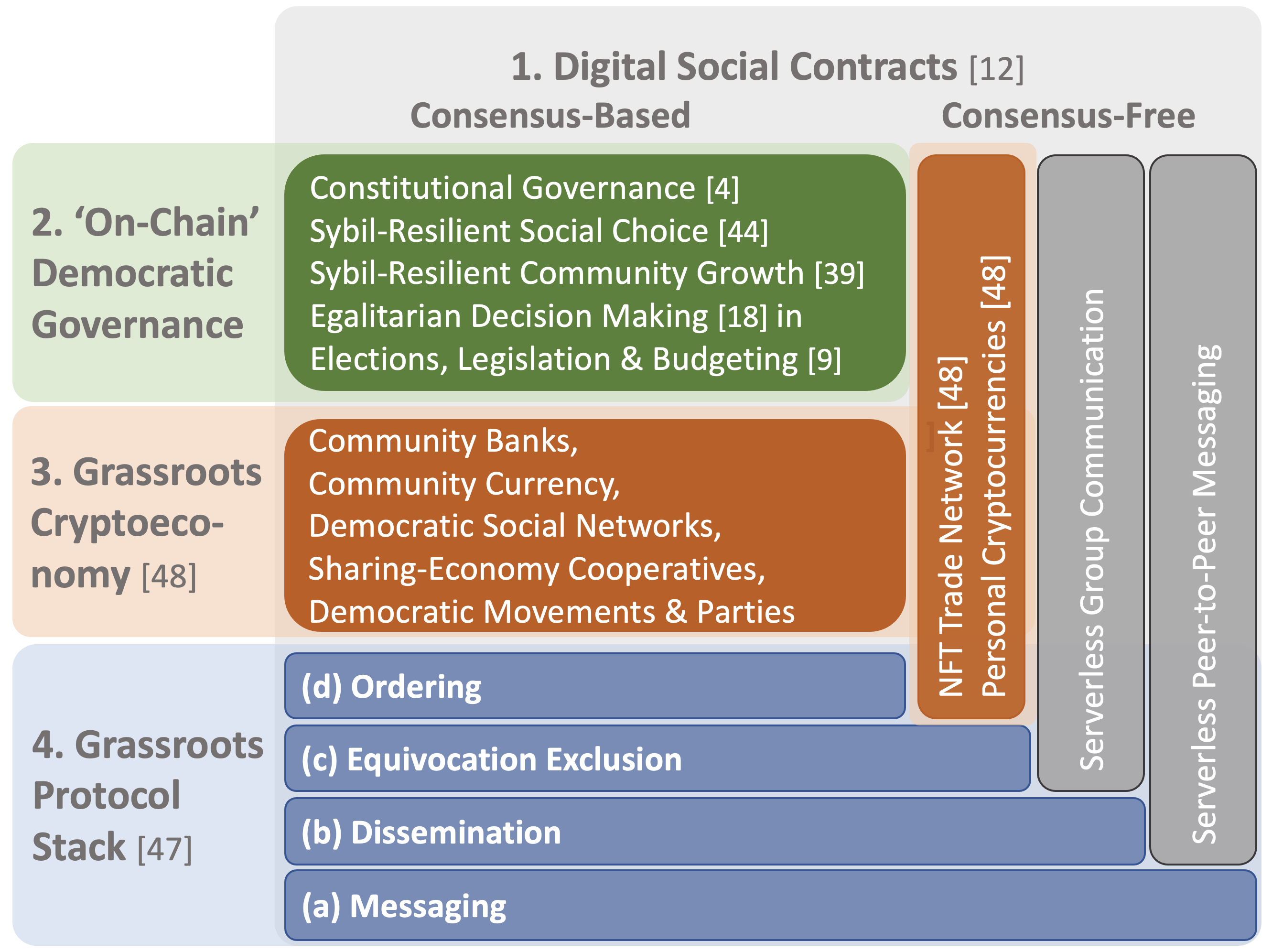}
\caption{Grassroots Democratic Metaverse Architecture}

\label{figure:metaverse}
\end{figure*}
\begin{enumerate}
\item
\textbf{Digital social contracts (Section \ref{section dsc})}  are Proudhon's~\cite{proudhon2004general} notion of the social contract---\emph{a voluntary agreement among free individuals}---ported to the digital realm~\cite{cardelli2020digital}.

\item
\textbf{Democratic Governance of DAOs  (Section \ref{section governance})}  considers the broad range of issues that need to be addressed by digital social contracts that instruct democratic governance, namely democratic DAOs~\cite{shapiro2018foundations,shahaf2019sybil, meir2020sybil, abramowitz2021beginning, bulteau2021aggregation, elkind2021united, button2019forking, abramowitz2021democratic}.  

\item \textbf{Grassroots cryptoeconomy (Section \ref{section cryptoeconomy})} aims to facilitate the capital-free bootstrap of grassroots cryptoeconomic communities, starting from self-sovereign personal cryptocurrencies~\cite{shapiro2022selfsovereign}.

\item
\textbf{Grassroots protocol stack (Section \ref{section stack})} supports the entire spectrum 
of communication and coordination needs of a grassroots democratic metaverse~\cite{shapiro2021multiagent},

\end{enumerate}


We expect this theoretical foundation and architecture to be realized as a common good, open and available to all. With it, autonomous democratic alternatives to existing digital autocracies and plutocracies may flourish.


\section{Digital Social Contracts}\label{section dsc}

Philosophically, a \emph{digital social contract} is  Proudhon's~\cite{proudhon2004general}  notion of the social contract---\emph{a voluntary agreement among free individuals}---ported to the digital realm~\cite{cardelli2020digital}. It is different from Rousseau's~\cite{rousseau1762social} notion, where membership in a nation bound by a social contract is not voluntary, and the nation is justified in using force to coerce its members.  In a digital social contract people agree to abide by the rules of the contract in their digital interactions with the other parties to the contract.  %
These rules are specified by a working code---as in standard communication protocols and in cryptocurrency-based smart contracts---so that the execution of a correct code by the people participating in the contract ensures that they abide by the contract.  Unlike standard smart contracts, the intended parties to a digital social contract are people, digitally identified,  not anonymous accounts or wallets.    Digital social contracts are truly autonomous and egalitarian in their execution: They are executed by the parties to the contract themselves, using their own computing devices (smartphones, at present), rather than by anonymous third parties, namely miners that charge a fee (gas) for their service.  The digital social contract may aim to exclude sybils with various  sybil-resilient mechanisms discussed in Section~\ref{section governance}).

Computationally, we distinguish two types of digital social contracts, depending on the depth of the protocol stack they employ (Section \ref{section stack}):  Consensus-based and consensus-free.  

A \emph{consensus-based digital social contract} is a smart contract autonomously executed by its vetted participants~\cite{cardelli2020digital,buterin2014next,de2021smart}.  It execution requires the full protocol stack presented in Section \ref{section stack}, including the ordering layer, as a correct execution of a smart contract requires consensus on the ordering of the acts of the parties to the contract.
A key application of smart contracts are DAOs~\cite{hassan2021decentralized,ethereum:dao}, which typically specify a form of plutocratic governance. Similarly, a key application of consensus-based digital social contracts are democratic DAOs (Section~\ref{section governance}).

A \emph{consensus-free digital social contract} is a voluntary agreement among free individuals that employs a consensus-free protocol, namely a protocol lower down the protocol stack.
In general, constitutional democratic vetting requires a consensus-based protocol. But in case of a digital social contract for  leader-based group, vetting of participants can be done by the group leader, similarly to standard server-based social networks, based on a simpler dissemination protocol.  Other consensus-free digital social contracts, e.g. the personal cryptocurrency network~\cite{shapiro2022selfsovereign} discussed in Section \ref{section cryptoeconomy}, are open, or permissionless, in that any person may participate.

We envision digital social contracts that could cut the umbilical cord that today connects the multitudes to surveillance capitalism-based services~\cite{zuboff2019age} (e.g. Facebook), autocratic `sharing-economy' apps (e.g. Uber, Airbnb), and plutocratic cryptoeconomy platforms (e.g. DeFi on mainstream cryptocurrencies), and supplant them with grassroots, peer-to-peer and community-based digital conduct -- social, economic, and political. This could be achieved via digital social contracts for a host of cryptoeconomic applications discussed in Section~\ref{section cryptoeconomy}.

\mypara{Research challenges}
%
%
\begin{itemize}

\item
Developing a programming language for digital social contracts.

\item Programming use-cases of digital social contracts.
\end{itemize}

\section{Democratic Governance of DAOs}\label{section governance}

We consider issues that need to be addressed for the coding of digital social contracts for democratic governance, namely democratic DAOs.  Democratic DAOs are related to today's DAOs~\cite{hassan2021decentralized,ethereum:dao}, but with important differences: Today's DAOs are usually among anonymous accounts, or wallets; their governance is typically plutocratic (one coin -- one vote; or one coin -- square-root vote, as in quadratic voting~\cite{lalley2018quadratic}); it is executed by anonymous third parties -- miners that require remuneration (gas); and it is executed on a predetermined platform and protocol. In contrast, a democratic DAO is among digitally identified individuals; its intended governance is democratic; it is executed by its participants; it is amenable to encompassing constitutional governance, as every aspect of the digital social contract, including the underlying protocol of execution as well as the democratic decision process itself, may be subject to democratic amendment.

The challenges of encoding a practical democratic DAO are immense, as it requires formalizing and integrating all aspects of digital democratic governance. These include: \emph{sybil-resilient social choice theory} that allows a  digital community to govern itself democratically despite a bounded penetration of sybils~\cite{shahaf2019sybil,meir2020sybil}; \emph{digital constitutional governance}, which provides for the constitutional democratic governance of a digital community, including the democratic amendment of the constitution itself~\cite{abramowitz2021beginning};
\emph{sybil-resilient community formation}, which provides for constitutionally-instated sybil-resilient community admission and expulsion protocols;  \emph{effective egalitarian decision making}, in which members are equal as proposers, discussants, coalition-builders and voters, addressing elections, legislation, and budgeting within a uniform framework~\cite{bulteau2021aggregation, elkind2021united};
\emph{democratic coalition formation, merging, and forking}, which includes  democratic procedures for deliberation and coalition formation within a community~\cite{elkind2021united}, and for deciding on the merging of communities and the forking of a community~\cite{button2019forking,abramowitz2021democratic}.

\mysub{`On-Chain' Constitutional Governance}\label{section constitutional}
Leading cryptocurrencies employ an informal governance process that often causes so-called \emph{hard forks}~\cite{webb2018fork}, in which an \emph{off-chain} social decision process results in the previous protocol being abandoned by some or all of the participants; and a new protocol is initialized, ideally in a state (e.g. account balances) that reflects the state of all participants in the abandoned protocol.

\emph{On-chain governance}~\cite{reijers2018now} aims to avoid hard forks by providing a decision process in which the cryptocurrency's protocol can be amended `from within' according to the rules of the protocol itself.

On-chain governance of cryptocurrencies is an open challenge~\cite{goodman2014tezos},
in particular as the governance shall not be limited only to the contract (e.g., as in Ethereum), but shall include governing the platform itself.
%
%
%
A recent work~\cite{abramowitz2021beginning} provides an initial axiomatization of relevant constitutional amendments and corresponding algorithmic methods, by considering a multiagent system of agents with preferences over the constitution.

\mypara{Research challenges}
\begin{itemize}

\item 
Developing the theory of democratic amendments of governance procedures, including the constitution itself.
  
\item 
Suggesting social choice mechanisms for on-chain forking, following recent work~\cite{forking}.

\end{itemize}

\mysub{Sybil-Resilient Social Choice Theory}\label{section two}
A cornerstone of the democratic metaverse is the ability to make joint community decisions. Viewing elections and voting as a standard tool for making joint community decisions, it is natural to look at computational social choice (COMSOC)~\cite{brandt2016handbook} for answers;
but keeping in mind that the problem of sybils poses a serious threat to any online community wishing to use some of these mechanisms:
  As a simple example, consider a community of $100$ agents, out of which $51$ are sybils (i.e., fake accounts);
  then, if a single binary decision is to be made -- for which the majority rule is a natural choice~\cite{may1952set} -- then the sybils effectively ``control'' the election. More generally, in any classical (non-sybil-resilient) voting scheme, a single sybil may tilt the decision or the elections result.

More concretely, a \emph{sybil-resilient election} consists of a community of $n$ agents, out of which at most a $\sigma$-fraction of the agents are sybils; and a \emph{sybil-resilient voting rule} gets a corresponding set of $n$ votes---without knowing which of the votes are from the sybils---and shall output an election winner, while satisfying some form of sybil safety.
COMSOC has yet to consider the problem of sybils in earnest for its methods and results to be applicable for a digital democracy; but there is some initial progress in that direction:
To define sybil-resiliency, Shahaf et al.~\cite{shahaf2018reality} and Meir et al.~\cite{meir2020sybil} follow reality-aware social choice~\cite{shapiro2017reality} in treating \emph{reality} (i.e., the status quo) as a distinguished alternative that must always be present. Based on this, \emph{sybil safety} requires that the status quo should change only if a majority among the non-sybils support the change; and \emph{sybil liveness} requires that the status quo can be changed even if all the sybils oppose the change. They show that supermajority voting rules---for various degrees of supermajority used---can satisfy both sybil safety and liveness, provided the fraction of sybils is below \nicefrac{1}{3}.
Slightly farther away, the works of Tran et al.~\cite{tran2009sybil} and Waggoner et al.~\cite{waggoner2012evaluating} (as well as others), can be viewed as showing some theoretical impossibility results regarding the possibility of voting rules that are resistent to sybils.

\mypara{Research Challenges}
%
%
\begin{itemize}

\item
Evaluating the theoretical lower bounds on the supermajority degree that guarantees certain safety-liveness tradeoffs practically.

\item
Studying sybil-resiliency for further COMSOC settings, such as multiwinner elections~\cite{mwchapter}, participatory budgeting~\cite{aziz2021participatory}, and constitutional amendments~\cite{abramowitz2021beginning,abramowitz2021amend}.


\end{itemize}



\mysub{Practical Egalitarian Decision Making}
Practical decision making mechanisms are critical for the success of democratic DAOs.


\mypara{Research Challenges}
\begin{itemize}

\item
Developing egalitarian decision making procedures, in which members are equal as proposers, discussants, negotiators, coalition-builders, and voters~\cite{elkind2021united}, addressing elections, legislation, and budgeting within a uniform framework~\cite{bulteau2021aggregation}, to facilitate egalitarian and effective democratic governance.
    
\item
Investigating democratic coalition formation, merging, and forking, that are democratic procedures for deliberation and coalition formation within a community~\cite{elkind2021united}, and for deciding on the merging of communities and the forking of a community~\cite{button2019forking,abramowitz2021democratic}.
    
\item
Developing unified, generic decision making processes that can encompass different COMSOC settings, similarly to metric-based aggregation methods, that, together with metric embeddings for these different social choice settings, may be useful~\cite{bulteau2021aggregation}), particularly if it can also be integrated with a uniform process for deliberation, voting, and ad-hoc coalition formation~\cite{elkind2021united,goel2016towards}. 


  

\end{itemize}

\mysub{Grassroots Community Formation}\label{section three}
%
%
Given adequate sybil-resilient voting rules, a digital community can tolerate some sybils, but not so if the fraction of sybils is too large (specifically above \nicefrac{1}{3}~\cite{shahaf2019sybil}). Considering the dynamic nature of community formation, the question arises how can a digital community grow while retaining a bounded-fraction of sybils? This relates to sybil identification~\cite{alvisi2013sok} and prevention~\cite{levine2006survey,siddarth2020watches}.

First, one has to distinguish between genuine personal identifiers and sybils~\cite{shahaf2020genuine}: 
Essentially, genuine personal identifiers are unique and singular, while while sybils are either duplicate or fake. Put differently, an honest person would only declare a single identifier as their genuine personal identifier (or declare a new identifier as a replacement, e.g. if their private key was lost or compromised), while a dishonest (`Byzantine') person would peddle sybils -- they would create multiple identifiers and claim each of them to be their own genuine personal identifier (duplicates); create identifiers and claim them to be genuine identifiers of people that do not exist (fake); or knowingly perpetuate sybils created by others.

The dynamics of community formation can be modeled via a transition-based system:
  Initializing with some community seed, the system transitions by adding or removing community members; while some upper bound on the fraction of sybils shall be maintained.
In particular, if an initial digital community consists of identifiers of honest people; if honest people tend to trust honest people more than they trust non-honest ones; and if the trust graph of a digital community---the graph depicting the trust relations in the community---is rich enough, then new people can be safely added to the community while maintaining a bound on the sybil penetration of the community. Intuitively, this can be done by ensuring sufficient connectivity in the trust graph. E.g., ensuring that the \emph{conductance} of the community's trust graph (which relates to the vertex and edge expansion of the graph) does not decrease following the addition of new (identifiers of) community members guarantees that the fraction of sybils that penetrate the community can be bounded~\cite{poupko2021building}.

\mypara{Research challenges}
%
%
\begin{itemize}
\item 
Proposing local connectivity measures---in contrast to graph conductance that is a global measure---that make it easier to understand what does a person need to do in order to be admitted to the community.  

\item
Performing extensive simulations to assess the quality of solutions for sybil-resilient community growth.

\item 
Developing mechanisms to encourage honest people to be judicious in granting their trust, encourage sybil hunting, and reward sybil detection; investigating the integration of a due process for sybil determination, and the cross-community interaction for strengthened sybil-resilience.

\end{itemize}

\section{Grassroots Cryptoeconomy}\label{section cryptoeconomy}

Mainstream cryptocurrencies---based on proof of work~\cite{bitcoin,ethereum} or stake~\cite{kiayias2017ouroboros}---grant participants power and wealth in accordance with their capital investment, thus benefiting the few and exacerbating economic inequality.  In contrast, the goal of grassroots cryptoeconomy~\cite{shapiro2022selfsovereign} is to bootstrap a thriving trust-based and community-based cryptoeconomy without external capital or credit and without reliance on third-party computing services.  
The proposed grassroots cryptoeconomic components include:

\mysubb{Self-sovereign personal cryptocurrencies}~\cite{shapiro2022selfsovereign} can form the basis of grassroots economies, with community cryptocurrencies~\cite{shahaf2021egalitarian} as an emergent phenomena.  A personal coin is best thought of as a transferable unit of credit issued by one person to another, and two people can establish a mutual line of credit between them~\cite{trustlines} simply by exchanging their personal coins.
    Liquidity in a grassroots economy can be achieved via such mutual credit lines, forsaking initial capital or external credit.  Fault-resilience and transaction-finality are provided by each agent regarding its own personal currency. 
    Self-sovereignty means that the responsibility for the economic and computational integrity of a personal currency resides with the person; as the value of a personal currency depends on such integrity, every person is incentivized to maintain them.

   \mysubb{Community banks/credit unions}~\cite{shapiro2022selfsovereign} may be realized as democratic DAOs, owned, operated and democratically-governed by their members.  Such a bank may streamline liquidity and simplify payments within a grassroots cryptoeconomic community that employs personal cryptocurrencies.  The bank issues its own currency and provides credit to its members at its discretion. It allows each member to deposit their personal coins and draw bank coins, up to the person's credit limit, in return for the members accepting its coins, which becomes a \emph{community currency}~\cite{circles-UBI,shahaf2021egalitarian,shapiro2022selfsovereign}. 
  The deposited personal coins are the bank's collateral, and any agent may redeem bank coins, first to its own personal coins held by the bank and any remaining balance to other personal coins held by the bank.
   A community currency can achieve sybil-proofness via a stake-based web of trust~\cite{shahaf2020genuine, gurin2020sybil}, being an essential ingredient of the community formation process discussed in Section~\ref{section governance}.  The process relies on trust among agents, and requires an incentive mechanisms that backs up trust edges by mutual sureties~\cite{gurin2020sybil}, denominated in the community currency redeemed by the community if the trusted personal identifier is determined to be a sybil (via a due process specified by the community's social contract).  This may require community members who wish to invite others to the community to lend them coins or provide them credit in this currency, to finance the mutual surety. 

    \mysubb{Universal Basic Income} (UBI). External support to a community bank---governmental or philanthropic---can boost the liquidity of the community and allow the bank to provide community members with a growing credit line in community currency, akin to UBI~\cite{shahaf2021egalitarian,trustlines,circles-UBI,howitt2019roadmap,assiagood,shapiro2022selfsovereign}.

    \mysubb{Democratic social networks}.  Commercial social networks follow the Feudal model of governance, where the corporate chief executive embodies all three branches of government  -- legislative, executive, and judicial.  Members have no civil rights and are meagerly-remunerated, if at all, for the exploitation of their digital capital (personal information) or labor (consuming ads and commercial content, creating engaging content). A democratic social network, realized as a democratic DAO,
     can be members-owned and operated, have proper democratic governance, and share revenues among members based on their contribution.

     \mysubb{Sharing-economy digital cooperatives}.  The ethos of earthly cooperatives includes autonomy and democratic conduct.  As such, democratic DAOs offer a natural embodiment for their digital counterparts: members' owned, operated and governed digital cooperatives.  In addition to digital cooperatives for social networks that offer a democratic and income-sharing alternative to Facebook,  we envision a true sharing-economy drivers-and-passengers-owned  cooperative that may offer a beneficial alternative to Uber, and a true sharing-economy owners-and-renters-owned cooperative that would be preferable to Airbnb.

\mypara{Research challenges}
\begin{itemize}

\item
Developing the concept of a network of community-issued currencies, concentrating on issues of liquidity and interoperability, both theoretically and practically, using multiagent simulations.

\item
Validating the notion of personal currencies and its viability as an infrastructure for sureties, theoretically and practically.

\end{itemize}

\section{Grassroots Protocol Stack}\label{section stack}

Executing the different applications enabled by digital social contracts (Section~\ref{section dsc}) on networked personal devices  requires a novel, alternative communication protocol stack.  Such a protocol stack would provide the entire spectrum of communication capabilities needed for the grassroots-formation of a democratic metaverse,  starting from peer-to-peer messaging and culminating in democratic DAOs. 
In contrast to a monolithic protocol (e.g. blockchain consensus~\cite{bitcoin}), a protocol stack can provide each application class the needed functionality in the most efficient, integrated, scalable way, as depicted in Figure \ref{figure:metaverse}.   Such a protocol stack  should
provides in an open and egalitarian way the entire spectrum of communication capabilities needed for the grassroots formation of a democratic metaverse~\cite{shapiro2021multiagent}, including:
\begin{enumerate}
    \item \emph{Messaging}. Serverless peer-to-peer communication, enabling serverless alternatives for personal messaging apps such as WhatsApp/Messenger/Telegram/Signal,  without control or surveillance by third parties.

    \item \emph{Dissemination}. Serverless group formation and communication; enabling serverless alternatives for groups messaging apps such as WhatsApp/Telegram/Facebook.
    \item \emph{Equivocation exclusion}. Serverless and consensus-free prevention of double-spending in NFT trade networks and personal cryptocurrencies~\cite{shapiro2022selfsovereign, collins2020online,guerraoui2018at2,naor2022payment}.
    \item \emph{Ordering}. An efficient, autonomous, and egalitarian distributed consensus protocol (e.g. ~\cite{keidar2021need,yin2019hotstuff,giridharan2022bullshark}) for the execution of consensus-based digital social contracts.
\end{enumerate}
An example of such a protocol stack is presented in reference~\cite{shapiro2021multiagent}.

\mypara{Research challenges}
\begin{itemize}

\item 
Considering various architectures for such a protocol stack and efficiently implementing the above-mentioned layers.

\item
Proposing practical, permissionless trust-based peer-to-peer messaging, dissemination and equivocation exclusion protocols.
\end{itemize}

\section{Outlook}

We have described a vision, an architecture, and  a roadmap for the grassroots formation of interconnected and interoperable autonomous digital communities, such as democratic social networks, sharing-economy digital cooperatives, democratic social movements, and political parties.
We have highlighted different subfields of MAS that are relevant to advance the vision and its different components.
Viewed as a whole, we discussed how the presented architecture and its components could enable the grassroots formation of a democratic metaverse.


\balance
\bibliographystyle{plain}
\bibliography{bib}

\end{document}